\documentclass{report}
 
\begin{document}

  
  \setlength{\textheight}{20cm}
  \setlength{\textwidth}{13.5cm}
  \renewcommand{\abstract}[1]{{ \footnotesize \noindent {\bf Abstract} #1 \\}}
  \renewcommand{\author}[1]{\subsubsection*{\it#1}}
  \newcommand{\address}[1]{\subsubsection*{\it#1}}
  
  %
  \def\ga{\mathrel{\mathchoice {\vcenter{\offinterlineskip\halign{\hfil
  $\displaystyle##$\hfil\cr>\cr\sim\cr}}}
  {\vcenter{\offinterlineskip\halign{\hfil$\textstyle##$\hfil\cr>\cr\sim\cr}}}
  {\vcenter{\offinterlineskip\halign{\hfil$\scriptstyle##$\hfil\cr>\cr\sim\cr}}}
  {\vcenter{\offinterlineskip\halign{\hfil$\scriptscriptstyle##$\hfil\cr>\cr\sim\cr}}}}}
 %
 %
%
\def\la{\mathrel{\mathchoice {\vcenter{\offinterlineskip\halign{\hfil
$\displaystyle##$\hfil\cr<\cr\sim\cr}}}
{\vcenter{\offinterlineskip\halign{\hfil$\textstyle##$\hfil\cr<\cr\sim\cr}}}
{\vcenter{\offinterlineskip\halign{\hfil$\scriptstyle##$\hfil\cr<\cr\sim\cr}}}
{\vcenter{\offinterlineskip\halign{\hfil$\scriptscriptstyle##$\hfil\cr<\cr\sim\cr}}}}}

 \chapter*{Cosmic inventory of energy densities: issues and concerns}
 \author{T.Padmanabhan}
 \address{IUCAA, PB 4, Ganeshkhind, Pune-411 007, India}
 \abstract{The dynamics of our universe is characterised by the  density parameters for cosmological constant ($\Omega_V$), nonbaryonic darkmatter($\Omega_{\rm wimp}$), radiation ($\Omega_R$) and baryons ($\Omega_B$). To these parameters --- which describe the smooth background universe --- one needs to add at least  another dimensionless number
 ($\sim 10^{-5}$)  characterising the strength of primordial fluctuations in the gravitational potential, in order to ensure formation of structures by gravitational instability.
  I discuss several issues related to the description of the universe in terms of these numbers
 and argue that we do not yet have a fundamental understanding of  these issues.   }
 
 \section{An inventory of energy densities}
 
 Based on dynamical considerations, we can divide the energy densities contributing to the 
 expansion of the universe into those due to (i) radiation and relativistic particles (with an equation of state $p=w\rho$
 and $w=1/3$), (ii) baryons ($w=0$), (iii) non relativisitic, non baryonic dark matter ($w=0$) and (iv) cosmological
 constant ($w\approx -1$). Current observations suggest that 
 the top two positions
 are held by cosmological constant ($\Omega_V \approx 0.6$) and non baryonic dark matter 
 ($\Omega_{\rm wimp} \approx 0.35$)  for which we have no 
 laboratory evidence. [The third position will  go to massive neutrinos if the recent observational
 results are confirmed.]
 The baryons in the universe contribute $\Omega_B \approx 0.02 h^{-2}$ which is about an order
 of magnitude larger than the baryons seen in the form of luminous matter,  indicating 
 that at least part of the baryons are dark. The energy density in electromagnetic radiation is dominated
 by CMBR and contributes about $\Omega_R \approx 2.5 \times 10^{-5} h^{-2}$. 
 
 These energy densities
 drive the expansion of a homogeneous and isotropic universe which is completely structureless. The 
 existence of small scale structures in the universe is believed to be due to gravitational instability
 which has amplified small initial fluctuations. Such a paradigm  requires the existence of
 nonzero fluctuations in the  gravitational
 potential energy due to primordial density fluctutations. Observations suggest that this can be characterised by a 
 dimensionless number ($\sim 10^{-5}$) in our universe. 
 
 I shall now try to describe some of the issues related to
 these numbers which requires investigation, focussing on the questions: Do we understand any of these components at a fundamental level or can we relate them to one another
 in a meaningful  way? Unfortunately, the answer today is `no'.

 \section{Cosmological constant: The theoretist's nightmare}
 
 Current observations  suggest that nearly 95 per cent of the matter in the universe is
 non baryonic. Cosmologists, over years, have learned to live with the existence of dark matter which 
 is structurally very different from the normal baryonic matter one is familiar with in the laboratory.
 But recent observations of supernova and CMBR seem to indicate that there are atleast
 {\it two} different components of dark matter, one made of weakly interacting massive particles
 contributing about $\Omega_{\rm wimp} \approx 0.35$ and another made of some form of 
 energy with negative pressure contributing about $\Omega_V \approx 0.6$. The simplest 
 choice for the latter is a cosmological constant with an equation of state $p=w\rho$
 where $w=-1$. Observations are, however, consistent with a somewhat broader range of 
 negative values for $w$ and it is possible to come up with more exotic equations of state
 (with negative pressure) for this component. If these observations do not go away 
 (and it looks unlikely that they will!), we have a serious theoretical problem in our hands. 
 
 Conventionally, there are two ways of interpreting the cosmological constant. In the first approach,
 one introduces $\Lambda$ as a parameter in Einstein's theory of gravity just as the newtonian constant
 $G$ is introduced as a parameter in this theory. In such a case, the action for matter coupled to gravity
 will have the form 
 \begin{equation}
 A = {c^3\over 16 \pi G} \int (R-2\Lambda) \, \sqrt{-g} d^4 x + \int {\cal L}_{\rm matter}\sqrt{-g} d^4 x 
 \label{eqn:action}
\end{equation}
where ${\cal L}_{\rm matter}$ is the matter lagrangian. In this approach, the numerical value of $\Lambda$ needs to be determined from observations, just like the
numerical value of $G$. In classical gravity, it is not possible to construct a dimensionless number
using the fundamental constants $\Lambda, G$ and $c$ which appear in the action; hence  the relative value of $\Lambda$ and 
$G$ does not make any sense and depends on the system of units. The situation, however, is different
once we add the planck constant $\hbar $ into the fray. (This does not require quantum gravity;
even in the description of quantum fields in a classical curved background, one naturally ends up
getting $\hbar$ in the analysis as in the case of, for example, particle creation by an expanding universe.)
 It is then possible to form the dimensionless number $\Lambda (8\pi G\hbar/c^3) \equiv \Lambda L_P^2 \approx
 10^{-120}$ where $L_P\approx 10^{-32} $  cm is the planck length. The fact that this dimensionless
 number --- made out of fundamental constants appearing in our theory ---  is so tiny, has been a source
 of mystery. If no other inputs are available, one would have expected such a small number to be actually
 equal to zero and one would look for a deep symmetry in the theory which requires vanishing of $\Lambda$;
 that is, the $\Lambda\sqrt{-g}$ term cannot exist in (\ref{eqn:action}) because of certain symmetry just as electromagnetic action cannot have a $m^2 A_i A^i$ term (making photon massive) because of gauge invariance. No such exact symmetry is known.
 
 There is, however, an alternative --- and more compelling --- way of interpreting the cosmological 
 constant which  make matters worse. In equation (\ref{eqn:action}) we can always add to the matter
 lagrangian  any arbitrary constant. A change ${\cal L}_{\rm matter} \to  {\cal L}_{\rm matter} -
 V_0$ is identical to the replacement $\Lambda \to \Lambda + 8\pi G V_0 / c^3$. 
 Since observations can only determine the net $\Lambda$, it 
 follows that we cannot just treat $\Lambda$ as a parameter in the theory; it takes contributions from the matter lagrangian and only the total value has observational consequence. After adding all the matter sector 
 contributions, the net result should finally satisfy the constraint 
 $\Lambda_{\rm net} L_P^2 \la 10^{-120}$. 
  As the universe evolves and cools some of the symmetries present at 
 high energies will be spontaneously broken. This could lead to a $V_0 \approx E^4$ where
 $E$ is the energy scale at which the symmetry is broken. For GUTs, $E\approx 10^{14}$ GeV and
 for electro-weak phase transition, $E\approx 100$ GeV. These contributions needs to be 
 precisely cancelled by an original $\Lambda $ term so that $\Lambda_{\rm net}$ is very tiny or zero.
 The only symmetry principle which we know
 that demands $\Lambda_{\rm net}$ to identically vanish is supersymmetry. But since supersymmetry
 is badly broken in nature, we would expect a contribution to $\Lambda_{\rm net}$ at the scale 
 $V_0 \approx E_{\rm susy}^4$ where $E_{\rm susy}$ is the energy scale at which supersymmetry is
 broken. This scale is still too high for comfort.
 
 All these difficulties related to the cosmological constant will plague us even if observations suggest
 that $\Lambda_{\rm net} =0$. 
 The cosmological observations suggesting that $\Lambda \approx H_0^{-2}$ adds a new layer
 of complications. Since the rate of expansion of the universe depends on the epoch while 
 $\Lambda_{net}$ is treated as a numerical constant in the action at low energies (below the electro weak scale, say), it is not clear why the relation 
 $\Lambda \approx H_0^2 $ should hold {\it around the present epoch}.  We now need to explain: (i)  why $\Lambda L_P^2$  is small but non zero and (ii) why is $\rho_\Lambda $ comparable
 to other dominant energy densities in the universe around the present epoch. 
 
 Ever since recent cosmological observations suggested the existence of a nonzero cosmological constant, there has been a flurry of theoretical activity to `` explain "
it, none of which even gets to the first base. One class of models invokes some version of anthropic principle; but since anthropic principle never predicted anything, I do not
consider it part of scientific methodology. The second class of models use a  scalar field with an ``appropriate" potential $V(\phi)$ to ``explain" the observations. In this case, we will take the $\Lambda$ term in (\ref{eqn:action}) to be zero [for unknown reasons] and will have a dynamically evolving cosmological ``constant" due to the
potential $V(\phi)$ in ${\cal L}_{\rm matter}$. These
models are all, however,  trivial and have no predictive power because it is always possible to choose a $V(\phi)$ to account for any sensible dynamical evolution of the universe. Since
the triviality of these models (which are variously called ``quintessence", ``dark energy" ....) does not seem to have been adequately emphasised in literature, let me briefly comment on this issue.

Consider any model for the universe with a {\it given}
$a(t)$ and some known forms of energy density $\rho_{\rm known} (t)$ (made of radiation,
matter etc) both of which are observationally determined. It can happen that this
pair does not satisfy the Friedmann equation for an $\Omega=1$ model. To be specific,
let us assume $\rho_{\rm known}<\rho_c$ which is substantially the situation in cosmology today. If we now want to make a consistent model of cosmology with $\Omega=1$, say,
we can invoke a scalar field with the potential $V(\phi)$. It is trivial to choose
$V(\phi)$ such that we can account for {\it any} sensible pair [$a(t), \rho_{\rm known} (t)] $ along the following lines: Using the given $a(t)$, we define two quantities
$ H(t)=(\dot a/a)$ and  $Q(t)\equiv 8\pi G \rho_{\rm known}(t) /3H^2(t)$. The required $V(\phi)$ is given parametrically by the equations:
 
\begin{equation}
V(t) = (1/16\pi G) H (1-Q)\left[6H + (2\dot H/H) - (\dot Q/1-Q)\right]
\end{equation}
 
\begin{equation} \phi (t) = \int dt \left[ H(1-Q)/8\pi G\right]^{1/2} \left[\dot Q/(1-Q )- (2\dot H/H)\right]^{1/2}
\end{equation}
All the potentials invoked in the literature are special cases of this formula. This result shows that 
{\it irrespective of what the future observations reveal about $a(t)$
and  $\rho_{\rm known}(t)$} one can always find a scalar field which will ``explain"
the observations. Hence this approach has no predictive power. What is worse, most of the $V(\phi)$
suggested in the literature have no sound particle physics basis and --- in fact ---
the quantum field theory for these potentials are very badly behaved on nonexistent.

It is  worth realising that the existence of a
 non zero cosmological constant will be a statement of fundamental significance and constitutes a conceptual contribution of cosmology to quantum gravity. The tendency of some cosmologists to treat $\Omega_\Lambda$
as one among a set of, say, 17 parameters [like $\Omega_{rad}, \Omega_B, n, ....$]
which needs to be fixed by observations, completely misses the point. Cosmological constant is special and its importance transcends cosmology.
Unfortunately,  we do not have at present a fundamental understanding of cosmological constant from any of the  approaches to quantum gravity. There are no nontrivial string theoretical models \cite{ewit01} incorporating $\rho_V>0$; loop gravity can incorporate it but does not throw any light
on its value. It should  be stressed that the nonzero value for
$\rho_V\neq 0$ does {\it not} imply deSitter (or even asymptotically deSitter)
spacetime. Hence the formalism should be capable of handling $\rho_V$ without deSitter geometry.
   
To give an example of a more fundamental way of thinking about cosmological constant, let me describe an idea in which cosmological constant is connected with the microstructure of spacetime. In this model we start with $\Lambda=0$ but generate a small value for this
parameter from two key ingredients:
 (i) discrete spacetime structure at Planck length and (ii) quantum gravitational
uncertainty principle. To do this, we first note that
cosmological constant can be thought of as a lagrange multiplier for proper volume
of spacetime in the action functional for gravity by rewriting the first term of (\ref{eqn:action}) as:
\begin{equation}
 \hbar^{-1} A_{\rm grav}={1\over 2L_P^2}\int d^4x R\sqrt{-g}-{\Lambda\over L_P^2}\int d^4x \sqrt{-g}; \label{eqn:xyz}
\end{equation}
In any quantum cosmological models which leads to large volumes for the universe, phase of
the wave function will pick up a factor of the form
$\Psi\propto \exp(-i(\Lambda/L_P^2){\cal V})$, where ${\cal V}$ is the four volume \cite{singhtp89}, from the second term in (\ref{eqn:xyz}).
Treating $(\Lambda/L_P^2,{\cal V})$ as conjugate variables $(q,p)$, we can invoke the standard uncertainty principle to predict
 $\Delta\Lambda\approx
L_P^2/\Delta{\cal V}$. Now we make the crucial assumption regarding the microscopic structure of the spacetime: Assume that there is a zero point length \cite{tp87} of the order of $L_P$
so that the volume of the universe is made of several cells, each of volume $L_P^4$. Then  ${\cal V}=NL_P^4$, implying a Poisson fluctuation $\Delta{\cal V}\approx
\sqrt{{\cal V}}L_P^2.$  and leading to 
\begin{equation}
\Delta\Lambda={L_P^2\over \Delta{\cal V}}={1\over\sqrt{{\cal V}}}\approx H_0^2
\end{equation}
which is exactly what cosmological observations imply!
Planck length cutoff (UV limit) and volume of the universe (IR limit) combine to give the correct $\Delta\Lambda$. 

After I gave this talk, I came to know that similar result was obtained earlier by Sorkin \cite{sorkin} based on a different model. The numerical result can of course arise in different contexts and it is probably worth discussing some of the conceptual components in my argument.  The first key idea is that, in this approach, $\Lambda$ is a stochastic variable with a zero mean and fluctuations. It is the rms fluctuation which is being observed in the cosmological context. This has two related implications: first, FRW equations now need to be sloved with a stochastic term on the right hand side and one should check whether the observations can still be explained.
Second, we may run into trouble if $\Lambda\approx H^{-2}$ at all epochs including the era of nucleosythesis since it {\it might} violate the bounds. Another key feature is that stochastic properties of $\Lambda$ need to be described by a quantum cosmological model. If the quantum state of the universe is expanded in terms of the eigenstates of some suitable opearator (which does not commute the total four volume operator), then one should be able to charecterise the fluctutations in each of these states. Generically, this will describe an ensemble of universes, with the fluctuations in cosmological constant inversly related to the size. Decoherence like arguments could then select out near classical, large volume universes.
 \cite{tp89}.
 
 While I am not optimistic about the {\it details} of the above model, I find it attractive to think of the observed cosmological constant as arising from quantum {\it fluctuations} of some energy density rather than from bulk energy density. This is relevant in the context of standard discussions of the contribution of 
 zero-point energies to cosmological constant. I would expect the correct theory to regularise the divergences and make the zero point energy finite and about $L_P^{-4}$ (see eg., \cite{dewitt64}, \cite{tp97}). This contributuion is most likely to modify the microscopic structure of spacetime (e.g if the spacetime is naively thought of as due to stacking of Planck scale volumes, this will modify the stacking or shapes of the volume elements) and will not affect the bulk gravitational field when measured at scales coarse grained over sizes much bigger than the Planck scales. In other words, large amounts of energy is soaked up by the quantum microstructure of the spacetime itself like a sponge soaking up water. This process, however, will leave a small residual fluctuation which will depend on the volume of the spacetime region which is probed. I would conjecture that the cosmological constant we measure corresponds to this residue. It is small, in the sense that it has been reduced from $L_P^{-4}$ to $L_P^{-4}(L_PH_0)^2$, which indicates the fact that fluctuations --- when measured over a large volume --- is small compared to the bulk value. It is the wetness of the sponge we notice, not the water content inside. 
 
 There are other ways of ``explaining" $\Lambda\approx H^{-2}$ and I will mention just two possibilities. If one assumes that every patch of the universe with size $L_P$  contained an energy $E_P$, then a universe with chasterctristic size $H_0^{-1}$ will contain the energy
 $E=(E_P/L_P)H_0^{-1}$. The corresponding energy {\it density} will be $\rho_V=(E/H_0^{-3})=(H_0/L_P)^2$ which is what one wants. The trouble, of course, is that we do not know why every length scale $L_P$ should contain an energy $E_P$.  Another possibility is to argue that the entropy of a patch of the universe containing vacuum energy density $\rho_V$is effectively the ratio between the Planck energy density $\rho_P=(E_P/L_P^3)$ and $\rho_V$; that is, $S\approx (E_P/L_P^3\rho_V)$. The idea being the most ordered state will have $\rho_V=\rho_P$ and lesser vacuum energy denisities will allow for larger number of configurations. If we now equate this entropy to that of the deSitter horizon $S=(H_0^{-2}/L_P^2)$, we immediately get $\rho_V=(H_0/L_P)^2$. Though very suggestive, it is based on two unjustified, non rigourous, conjectures. I mention these possibilities only because I believe the result $\rho_V=(H_0/L_P)^2$ will eventually arise from some deep geometrical feature rather than, say, from the slow roll over of a scalar field, introduced for this purpose.

 \section{ Energy in primordial gravitational potential fluctuations}
 
 A strictly homogeneous and isotropic universe cannot develop significant amount of structure over the timescale $(G\rho)^{-1/2}$. The very fact that our universe has structures shows that there must have been small deviations from homogeneity in the energy density at earlier phases. The gravitational potential $\phi(t,{\bf x})$ due to fluctutations in the energy density $\delta\rho(t,{\bf x})$ satisfies an equation of the form $\nabla^2\phi
 \propto (\delta/a)$ in the relevant scales. Taking $\delta\rho$ to be a gaussian at sufficiently early epochs, it follows that the fluctuations in the gravitational potential can be charecterised by the power per logarithmic band in the suitably defined ${\bf k}$ space:
 \begin {equation}
 \Delta^2_\phi(t,k)={k^3 P_\phi(t,k)\over 2\pi^2}\propto {P_\delta(t,k)\over k}
 \end {equation}
 where $P_\phi =|\phi_k|^2 $ and $P_\delta = |\delta_k|^2$ are the power spectra of gravitational potential and density
 contrast. 
 If this power is strongly scale dependent then we will either have copious production of small scale objects and blackholes (if the power at small scales dominate) or we will have stronger and stronger devitaions from FRW universe at larger and larger scales (if the power at large scales dominate). To allow for the kind of universe we see, it is important that $\Delta^2_\phi(t,k)$ is (at least, approximately) scale invariant, requiring $P_\delta(t,k)=A k$ with some constant $A$ \cite{hz70}. It follows that $\Delta^2_\phi(t,k)=\Delta^2_\phi(t)$; further at large scales, the evolution of the universe can be described by linear theory, and in a $\Omega=1$ model, $\Delta^2_\phi(t)$ does not evolve with time. We thus reach the remarkable conclusion that our universe is charecterised by a dimensionless number $\Delta^2_\phi\propto AH_0^4$. This number represents the strength of primordial gravitational fluctuations, or --- equivalently --- the strength of gravitational power at large scales. Observations of CMBR \cite{tpdna92} suggest that this number is about $10^{-5}$.
 
 Considering the fact that all of structure formation is essentially a transfer of this power from large scales to small scales, it is surprising that this number has received so little attention in the literature. It is easy to argue that this number could not have been more than a factor 100 off in either direction and still  lead to the kind of universe we live in. But we have no real  clue as to why it is about $10^{-5}$ ! If the fluctuations were generated by inflation, then this number is related to the parameters in the lagrangian for the inflaton field. For natural values of these parameters, one gets totally wrong answers. (This is in spite of grandoise claims as to how CMBR verifies inflation; if we take an initial spectrum $P=Ak^n$ with two parameters $A$ and $n$, any scale invariant mechanism will get $n=1$ and the {\it real}  test of the theory is to predict $A$ - a test  which inflation flunks). The model based on cosmic strings, in contrast, does get this number right as the dimensionless ratio $G\mu=(E_{\rm string}/E_P)^2$;  but, of course, this model has other problems.
 
 \section{Energy density of radiation}
 
 One of the most precisely measured energy densities is that due to cosmic microwave background radiation
 which dominates the radiation background in all wavebands. Today this energy density contributes
 $\Omega_R \approx 2.4 \times 10^{-5} h^{-2} (T/2.73\, {\rm K})^4$. The numerical value is probably not of
 much significance since it changes with the epoch as $(1+z)^4$. It is, however, possible to construct
 a conserved dimensionless number $N = a_0^3 n_R$ which represents
 the total number of photons (or the entropy of the radiation field) in the universe. Taking
 $a_0 = H_0^{-1} |\Omega_0 -1|^{-1/2}$, it follows that $N \ga  3 \times 10^{86} h^{-3}$.  [Note that, if $\Omega_0$ is 
 driven close to unity by inflation, then $\Omega_0-1$ could be contributed by the gravitational
 wave background generated in the same process.] This number, again, has no simple interpretation.
 In particular, if we assume that the universe had Planck temperature $T_P\approx 10^{19}$ GeV
 when its volume was about $L_P^3$, then the current CMBR temperature should be about
 $10^{-29}$ K! Obviously, some physical process should have increased the entropy of the 
 universe by a large factor (about $10^{90}$) in order to keep the universe warm enough today.
 Until this process is identified and pinned down, we must accept that we do not understand the 
 amount of energy density present in radiation today.  
 
 The above analysis is clearly related to the numerical value of the quantity $a(t) T_R(t)$ in our universe.
 If we assume that different particle species including the wimps were in equilibrium with radiation at
 sufficiently early epoch, then we can relate the number density of any non baryonic particle species to the 
 number density of photons. Given the masses and interaction strength of the wimp, it is then straight
 forward to calculate $\Omega_{\rm wimp}$ in terms of $\Omega_R$. (In the case of massless neutrinos, 
 for example, this is a fairly standard text book exercise but the principle is the same for any other particle
 species.) 
 Thus the computation of $\Omega_{\rm wimp}$ is related to our understanding of the particle physics 
 model {\it plus} our understanding of $\Omega_R$. 
 
 \section{Baryonic energy density}
 
 The most natural scenario for the universe would start with equal number of 
 baryons and anti baryons leading to $\Omega_B = 0$ today. The fact that we have a tiny number
 density of net baryons with a photon to baryon ratio being $n_B/n_R \approx 2.7 \times 10^{-8} \Omega_Bh^2 \simeq 5.4 \times 10^{-10}$ is yet
 another mystery in the current universe. (This is discussed in greater detail
 by Subir in his talk.) 
 
 Since bulk of the luminous matter is made of baryons, it is of course possible to estimate the 
 value of $\Omega_B$ at different epochs. In particular, the estimate $\Omega_Bh^2 \approx 0.02$
 from the big bang nucleosynthesis seems to be consistent with the recent BOOMERANG data which --- after 
 some initial hiccups (see eg., \cite{tpshiv01}) --- now gives $\Omega_Bh^2 = 0.02 \pm 0.005$. 
 The  MAXIMA data, however, still gives $\Omega_Bh^2 =0.05 \pm 0.014$ which is on the higher side.
 It is, however, not clear whether these results are 
 consistent with the measurement of baryonic density in the IGM. A careful study
 of Lyman-$\alpha$ absorbers in the IGM \cite{tirthantp} suggest that $\Omega_Bh^2 \ga 0.022 (\Gamma_{12}/1.2)^{1/2}$ where
 $\Gamma$ is the photoionization rate due to 
  the meta galactic ionizing
 flux in units of $10^{-12}s^{-1}$.  Since one believes that $\Gamma  \ga 1.5$, this result suggests a baryonic density in 
 IGM which is marginally inconsistent with the BBN. Only further work can show whether
 this result is due to inadequacies in modelling or whether it represents a genuine difficulty.
 
 \bigskip
 
 \noindent {\Large{\bf Acknowledgement}}
 
\noindent I thank K. Subramanian for useful discussions.

  \end{document}